\documentstyle[12pt]{article}
\newcommand{\reell}
{\kern+.23em\sf{1}\kern-.61em\sf{1}\kern+.76em\kern-.25em}

\begin{document}
\begin{flushright}
{\footnotesize DFTT 22/95 } \\
\ \hfill {\footnotesize CBPF-NF-17/95 }
\end{flushright}




\begin{center}
{\large\bf The Role of the Central Element in the Quantum Algebra
Underlying the Twisted XXZ Chain}

\ \\[2mm]

by

 \ \\[2mm]

{\sl M.R-Monteiro$^{\dagger ,\Delta ,a}$, I. Roditi$^{\Delta ,b}$,
L.M.C.S. Rodrigues$^{\Delta ,c}$},
and {\sl S. Sciuto$^{\dagger, d}$}\\[0.5cm]

$^{\dagger}$Dipartimento di Fisica Teorica dell'Universit\'a di
Torino
\\[-1mm]
and Sezione di Torino dell'INFN, Via Pietro
Giuria 1, \\[-1mm]
10125, Torino, Italy\\[2mm]

$^{\Delta}$Centro Brasileiro de Pesquisas
F\'{\i}sicas - CBPF\\[-1mm]
Rua Dr. Xavier Sigaud, 150\\[-1mm]
22290-180, Rio de Janeiro, RJ, Brasil\\[2mm]
\end{center}

\vspace{8cm}
\noindent
\rule{7cm}{0.2mm}\\
{\footnotesize e-mail:}\\
{\footnotesize (a) mmont@CBPFSU1.CAT.CBPF.BR}\\
{\footnotesize (b) Roditi@CBPFSU1.CAT.CBPF.BR}\\
{\footnotesize (c) L\'{\i}gia@CBPFSU1.CAT.CBPF.BR}\\
{\small (d) Sciuto@TO.INFN.IT}

\newpage
{}~
\vspace{5cm}
\begin{center}
{\sc Abstract}
\end{center}

We study the relationship among the XXZ Heisenberg model and
three models obtained from it by various transformations.
In particular, we emphasize the role of a non trivial central
element $t^Z$ in the underlying algebra and its relationship
with the twisted boundary conditions,
$S^{\pm}_{N+1}=t^{\pm N}S^{\pm}_1$.\\[0.5cm]

\noindent
{\bf Key-words:} \hfill\parbox[t]{13.5cm}{Quantum Groups; Quantum
Spin Chains; Integrable Models.} \\[2mm]

\newpage

The purpose of this letter is to study the relationship among three
XXZ-like models: two of them are obtained from the periodic
spin-$1/2$ XXZ chain \cite{vega} by performing De Vega
\cite{re} and Reshetikhin \cite{li} like transformations in the $L$
and $R$ matrices such that the transformed matrices still obey the
Yang-Baxter \cite{bax,yan} and the Fundamental Commutation Relations
(FCR) [6-8] and the third one is the XXZ-like model studied in
\cite{ku}. We find that the Hamiltonian with twisted boundary
conditions comes out when the algebra contains a non-trivial central
element, independently of the presence of the second parameter of
deformation in the commutation relations.

We start briefly reviewing a few well-known elements of the Quantum
Inverse
Scattering Method (QISM) [6-8] applied to the periodic quantum
spin-$1/2$ XXZ chain. The Hamiltonian describes a system of $N$
nearest
neighbour interacting particles of spin-$1/2$,
\begin{equation}
H=\sum^{N}_{\ell =1}(S^+_{\ell}S^-_{\ell +1}+S^-_{\ell}
S^+_{\ell +1}+\Delta S^3_{\ell}S^3_{\ell +1}) ,
\ \ \ \ \ \ \ \ ({\vec{S}}_{N+1}\equiv {\vec{S}}_1)\ ,
\end{equation}
with $S^{\pm}=S^1\pm iS^2$, $\vec{S}=\frac{1}{2}\ \vec{\sigma}$ and
$\vec{\sigma}=(\sigma^1,\sigma^2,\sigma^3)$ the Pauli matrices, and
$\Delta$
the anisotropy. $H$ acts on the $2^N$-dimensional Hilbert space,
\begin{equation}
{\cal H}=\prod^{N}_{\ell =1}\ \otimes  \ \eta_{\ell}\ ,
\end{equation}
where $\eta_{\ell}\equiv C\!\!\!\!I^{~2}$, from now on called the
internal space of the site ``$\ell $''.

In the QISM one introduces an auxiliary problem with the help of the
so-called Lax operator (belonging to End $(V\otimes \eta_{\ell})$
with $V\equiv C\!\!\!\!I^{~2}$ the auxiliary space) which in this
model is
\begin{equation}
L_{\ell}(\lambda )=\left(
\begin{array}{ll}
sh[\gamma (\lambda +S^3_{\ell})] & i \sin \gamma \ S^-_{\ell}\\
&\\
i\sin \gamma \ S^+_{\ell} & sh[\gamma (\lambda -iS^3_{\ell})]
\end{array}
\right)\ ;
\end{equation}
$\lambda$ is a complex number called  spectral parameter and
$\gamma$ is a parameter associated to the anisotropy, $\Delta$,
of the
Hamiltonian according to
\begin{equation}
\Delta =\frac{q+q^{-1}}{2}
\end{equation}
where $q=e^{i\gamma}$.

Associated to $L_{\ell}(\lambda )$ one introduces an invertible
matrix
$R(\lambda )$ belonging to End $(V\otimes V)$ which in this case
is
\begin{equation}
R(\lambda )=\left(
\begin{array}{cccc}
a(\lambda ) & 0 & 0 & 0\\
0 & c(\lambda ) & b(\lambda ) & 0\\
0 & b(\lambda ) & c(\lambda ) & 0\\
0 & 0 & 0 & a(\lambda )
\end{array}
\right)
\end{equation}
where
\begin{equation}
a(\lambda )=sh[\gamma (\lambda +i)]\ , \ b(\lambda )=i \sin \gamma \
, \
c(\lambda )=sh \gamma \lambda \ .
\end{equation}
$R(\lambda )$ satisfies the Yang-Baxter equation
\begin{equation}
R_{12}(\lambda_{12})R_{13}(\lambda_{13})R_{23}(\lambda_{23})=
R_{23}(\lambda_{23})R_{13}(\lambda_{13})R_{12}(\lambda_{12})
\end{equation}
with $\lambda_{ij}=\lambda_i-\lambda_j$, and $L_{\ell}(\lambda )$
obeys the FCR
\begin{equation}
R_{12}(\lambda_{12})L_{1\ell}(\lambda_1)L_{2\ell}(\lambda_2)=
L_{2\ell}(\lambda_2)L_{1\ell}(\lambda_1)R_{12}(\lambda_{12})\ .
\end{equation}
In (7) and (8)
\begin{equation}
R_{12}\equiv \sum_{i}\ a_i \otimes b_i\otimes {\reell} \ , \
R_{13}\equiv \sum_{i}\ a_i\otimes {\reell}\otimes b_i\ , \
R_{23}\equiv \sum_{i}{\reell} \otimes a_i\otimes b_i
\end{equation}
with the $R$ matrix written as
\begin{equation}
R(\lambda )=\sum_{i} \ a_i\otimes b_i
\end{equation}
and the additional indices, 1 and 2, in the $L_{\ell}$ matrix in (8)
follow
\begin{equation}
L_{1\ell}\equiv L_{\ell}\otimes {\reell}\ \ \ , \ \ \
L_{2\ell}\equiv {\reell} \otimes L_{\ell}\ .
\end{equation}
We notice that comparing (7) and (8) for the case under
consideration
((3), (5) and (6)), one has
\begin{equation}
L_{1\ell}(\lambda )=R_{1\ell}(\lambda -i/2)\ ,
\end{equation}
where the third auxiliary space is identified as the internal
``$\ell$'' space.

For the vertex-type models that we are considering, a local
Hamiltonian can
be obtained as:
\begin{equation}
H=\sum^{N}_{\ell =1}\ H_{\ell ,\ell +1}=
\left. C \ \frac{d\ln \tau}{d\lambda}\right|_{\lambda =i/2}\ ,
\end{equation}
with $C$ a constant and
\begin{equation}
\tau =Tr_0(L_{0N}L_{0N-1}\cdots L_{01})\ ,
\end{equation}
where the trace is over the auxiliary space denoted by ``$0$''.
Thanks to the property (12) and to $R(0)=P$, with $P$ the
permutation
operator in the tensor product space under consideration,
one can write
\begin{equation}
H_{\ell ,\ell +1}=\left. C\  \frac{d}
{d\lambda}\ (PR)_{\ell ,\ell +1}
\right|_{\lambda =0}\
\end{equation}
and with $R$ given in (5), one gets the Hamiltonian (1),
apart from
an additive constant.

In order to study the algebraic structure, after a similarity
transformation on (7) and (8) one can factor out the $\lambda$
dependence
in $R$ and $L$ \cite{fil}
\begin{eqnarray}
L_{\ell}(\lambda ) &=& \frac{1}{2}\ (q^{-i\lambda }
L^+-q^{i\lambda}
L^-)\\
R(\lambda ) &=& \frac{1}{2}\
(q^{-i\lambda}R-q^{i\lambda}R')\ ,\nonumber
\end{eqnarray}
where
\begin{equation}
L^+=\left(
\begin{array}{ll}
q^{S^3} & \Omega S^-\\
0 & q^{-S^3}
\end{array}
\right)\ \ \ ,
\ \ \ L^-=\left(
\begin{array}{rr}
q^{-S^3} & 0\\
-\Omega S^+ & q^{S^3}
\end{array}
\right)
\end{equation}
and
\begin{equation}
R=\left(
\begin{array}{llll}
q & 0 & 0 & 0\\
0 & 1 & \Omega & 0\\
0 & 0 & 1 & 0\\
0 & 0 & 0 & q
\end{array}
\right)\ \ \ ,
\ \ \ R'=\left(
\begin{array}{llll}
q^{-1} & 0 & 0 & 0\\
0 & 1 & 0 & 0\\
0 &-\Omega & 1 & 0\\
0 & 0 & 0 & q^{-1}
\end{array}
\right)
\end{equation}
with $\Omega =q-q^{-1}$, and $R$, $R'$ satisfying
\begin{eqnarray}
&PRPR'={\reell}&\ ,\\
&R-R'=\Omega P&\ ,\nonumber
\end{eqnarray}
$P$ being the permutation operator on the tensor
product space
$V\otimes V$.

Substituting (16) in the FCR one gets seven equations; it
can be shown that due to (19) only three of them are
independent
and they  can be chosen as:
\begin{eqnarray}
R_{12}L^{\varepsilon}_{1\ell}L^{\varepsilon}_{2\ell}&=&
L^{\varepsilon}_{2\ell}L^{\varepsilon}_{1\ell}
R_{12}\ \ \ \ \ \
(\varepsilon =\pm ) \\
R_{12}L^+_{1\ell}L^-_{2\ell}&=&
L^-_{2\ell}L^+_{1\ell}R_{12}\ . \nonumber
\end{eqnarray}
Let us forget for a moment that the $\vec{S}$ are
related  to the Pauli
matrices and consider them  as abstract elements of
an algebra; then
equations (20) reduce to
\begin{eqnarray}
{[}S^3,S^{\pm}] &=& \pm S^{\pm} \\
{[}S^+,S^-] &=& \frac{q^{2S^3}-q^{-2S^3}}{q-q^{-1}}\equiv
{[}2S^3]_q\ .\nonumber
\end{eqnarray}
which are the commutation relations of the quantum algebra
$SL_q(2)$
\cite{drin,la}. It is obvious that equations (21) are trivially
satisfied in the spin-$1/2$ representation of $SL(2)$ since
in this case
$S^3=0,\pm 1/2$ and $[n]_q=n$ for $n=0,\pm 1$, with
$[x]_q\equiv \frac{q^x-q^{-x}}{q-q^{-1}}$.

The co-structure is obtained by analysing the product for
different
internal spaces of two $L^{\varepsilon}$
operators (as before, the matrices  $\vec{S}$ are again
taken as
arbitrary elements of
an algebra) defined in the same auxiliary space denoted
by ``$0$'',
\begin{equation}
\Delta (L^{\pm})\equiv L^{\pm}_{0m}\
L^{\pm}_{0\ell}\ ,
\end{equation}
which reduces to
\begin{eqnarray}
\Delta q^{\pm S^3} &=& q^{\pm S^3}\otimes q^{\pm S^3}\\
\Delta S^{\pm} &=& q^{S^3}\otimes S^{\pm}+S^{\pm}\otimes
q^{-S^3}
\ .\nonumber
\end{eqnarray}
Since $\Delta (L^{\pm})$ satisfies the $FCR$ (eq.20),
$\Delta q^{\pm S^3}$
and $\Delta S^{\pm}$ given in (23) satisfy (21)
non-trivially.

The co-product defined in (22) is non co-commutative, i.e.,
\begin{equation}
\Delta '(L^{\pm}) \equiv L^{\pm}_{0\ell}L^{\pm}_{0m}\neq
\Delta (L^{\pm})
\end{equation}
but one can show that there is an invertible matrix
$\tilde{R}\in$  End
$(\eta_{\ell}\otimes \eta_m)$ such that
\begin{equation}
\tilde{R}_{m\ell}L^{\pm}_{0m}L^{\pm}_{0\ell}=
L^{\pm}_{0\ell}L^{\pm}_{0m}\tilde{R}_{m\ell}\ .
\end{equation}
Using $L^+_{21}(q)=L^-_{12}(q^{-1})$, which holds in
the spin-1/2
representation, $R_{21}(q)=R'_{12}(q^{-1})$ and (19),
one easily checks
that
\begin{equation}
\tilde{R}_{m\ell}=PR_{m\ell}P\ ,
\end{equation}
with $P$ the permutation operator in the tensor
product space
$\eta_{\ell}\otimes \eta_m$. Defining
$\tilde{R}(\lambda )=\frac{1}{2}\
(q^{-i\lambda}\tilde{R}-q^{i\lambda}\tilde{R}')$ with
$P\tilde{R}P\tilde{R}'={\reell}$, one easily
checks that  $ \tilde{R}(\lambda )=P\tilde{R}(\lambda )P$
and that the Yang-Baxter
equation (7) can be rewritten as:
\begin{equation}
\tilde{R}_{12}(\lambda_{12})\tilde{R}_{13}(\lambda _{13})
\tilde{R}_{23}(\lambda_{23})=
\tilde{R}_{23}(\lambda_{23})\tilde{R}_{13}(\lambda _{13})
\tilde{R}_{12}(\lambda_{12})\ .
\end{equation}

In the case under consideration ((5-6)),
$\tilde{R}(\lambda )=R(\lambda )$.

We are now going  to analyse further deformed structures
starting from
the XXZ spin-$1/2$ chain we have just presented. A new complex
parameter $t$ is introduced through the transformations:
\begin{equation}
g_1=t^{S^3\otimes Z}\ \ , \ \ g_2=t^{Z\otimes S^3}
\end{equation}
where $Z$ is a new operator; when acting on the
auxiliary
space or on the physical space of a single site, the algebra
generated by
$(\vec{S},Z)$ will be taken in its fundamental representation;
thus,
unless stated otherwise,
\begin{equation}
Z=\frac{1}{2}\ {\reell} \ \ , \ \ \vec{S}=
\frac{1}{2}\ \vec{\sigma}\ .
\end{equation}
In this case, it is easy to check that
\begin{equation}
[R_{12}(\lambda ),\ g_1g_2]=0\ .
\end{equation}
Using $g_1$ and $g_2$ we also define
\begin{equation}
F_{12}=g_1g^{-1}_2=t^{S^3\otimes Z-Z\otimes S^3}
\end{equation}
which satisfies
\begin{eqnarray}
PF_{12}PF_{12} &=& {\reell} \\
F_{12}F_{13}F_{23} &=& F_{23}F_{13}F_{12}\nonumber
\end{eqnarray}

The XXZ chain will be denoted by case A. Using
$R(\lambda ),\ g$
and $F$ and with the help of De Vega \cite{re} and
Reshetikhin \cite{li} type transformations we
shall build  other
three possible cases:

\vspace{1mm}
\noindent
{\bf CASE A:}
\begin{equation}
R^A(\lambda )=R(\lambda )\ ; \
L^A_{1\ell }(\lambda )=R^A_{1\ell}(\lambda -i/2)\ ; \
\tilde{R}^A(\lambda )=PR^A(\lambda )P\ ;
\end{equation}

\vspace{1mm}
\noindent
{\bf CASE B:}
\begin{equation}
R^B(\lambda )=R^A(\lambda )\ ; \
L^B_{1\ell }(\lambda )=g_1L^A_{1\ell}(\lambda )g_1\ ; \
\tilde{R}^B(\lambda )=F^{-1}\tilde{R}^A(\lambda )F^{-1}\ ;
\end{equation}

\vspace{1mm}
\noindent
{\bf CASE C:}
\begin{equation}
R^C(\lambda )=FR^A(\lambda )F\ ; \ L^C_{1\ell}(\lambda )=
g^{-1}_{\ell}L^A_{1\ell}g^{-1}_{\ell}\ ; \
\tilde{R}^C(\lambda )=\tilde{R}^A(\lambda )\ ;
\end{equation}

\vspace{1mm}
\noindent
{\bf CASE D:}
\begin{equation}
R^D(\lambda )=R^C(\lambda )\ ; \ L^D_{1\ell}(\lambda )=
R^D_{1\ell}(\lambda -i/2)=FL^A_{1\ell}(\lambda )F\ ; \
\tilde{R}^D(\lambda )=\tilde{R}^B(\lambda )\ ;
\end{equation}
we notice that for $g_{\ell}$ in (35) we take as
auxiliary space the
first vectorial space and as the ``$\ell$'' internal
space the second
vectorial space. Thus $g_{\ell}\equiv g_2$ where $g_2$
is defined in
(28) with the second vectorial space taken as the
internal ``$\ell$'' space.

As  previously discussed for the case A, it can
easily be checked for
the cases B,C and D, using (30), that
$L^{\bullet},\ R^{\bullet}$ and
$\tilde{R}^{\bullet}(\bullet =$A,B,C or D) satisfy
(7), (8), (25),
(27) and looking
at eqs. (33-36) one immediately realizes that
$R^{\bullet}(0)=P$, since $F$
satisfies (32).

The case B is an example of the procedure described
in \cite{re}
since it preserves the $R$ matrix, transforming
the $L$ matrix. The
matrix $\tilde{R}^B$ is related to $\tilde{R}^A$
by a Reshetikhin
transformation \cite{li}; thus the de Vega
transformation on the
matrix $L$ can be seen as an
application of the Reshetikhin theorem at
the level of QISM.

Instead the case C is got from the case A by
the dual of the
transformation sending A into B; we shall call
the latter a ``de Vega
transformation''
and the former a ``dual Reshetikhin transformation.''
These two
transformations are related to
each other by the interchange of the role of the
auxiliary space and
of the site space; this is particularly clear if
one compares
$L^B_{1\ell}=g_1L^A_{1\ell}g_1$ with
$L^C_{1\ell}=g^{-1}_{\ell}
L^A_1g^{-1}_{\ell}$.

Noticing that (34-36) imply
\begin{eqnarray}
L^D_{1\ell} &=
& g^{-1}_{\ell}L^B_{1\ell}g^{-1}_{\ell}\ , \\
L^D_{1\ell} &=& g_1L^C_{1\ell}g_1\ , \nonumber
\end{eqnarray}
one realizes that the case D \cite{ku} can be
obtained either from B
by applying a dual Reshetikhin transformation
or from C through a de
Vega transformation.

We are now going  to show that the cases connected
by a dual
Reshetikhin transformation have the same Hamiltonian.
In cases C
and D using (14), (35) and (37) we get
\begin{eqnarray}
\tau^C &=& G^{-1}\tau^AG^{-1}  \\
\tau^D &=& G^{-1}\tau^BG^{-1} \nonumber
\end{eqnarray}
with
\begin{equation}
G=\prod^{N}_{\ell =1}\ g_{\ell}=
t^{\frac{1}{2}S^3_{total}}\ .
\end{equation}
As
\begin{equation}
H=C\tau^{-1}\ \left.\frac{d\tau}
{d\lambda}\right|_{\lambda =i/2}\ ,
\end{equation}
using $[g_{\ell}g_{\ell +1},R_{\ell ,\ell +1}]=0$,
which in the case
we are considering means that $H^A_{\ell ,\ell +1}$
is invariant
under a simultaneous rotation of the sites $\ell$
and $\ell +1$
around the z-axis, we obtain
\begin{eqnarray}
H^C &=& GH^AG^{-1}=H^A \\
H^D &=& GH^BG^{-1}=H^B \ . \nonumber
\end{eqnarray}
The Hamiltonian can be directly written in terms
of $R_{\ell ,\ell
+1}$ (eq. 15) only in the cases A and D, where
the identity
$L_{1\ell}(\lambda )=R_{1\ell}(\lambda -i/2)$ holds.
However, it is
amusing to observe that, since $PR^A=
\tilde{R}^AP=\tilde{R}^CP$ and
$PR^D=\tilde{R}^DP=\tilde{R}^BP$, in all cases we
can write the
Hamiltonian as
\begin{equation}
H^{\bullet}=
\sum_{\ell =1}^{N}\ H^{\bullet}_{\ell ,\ell +1}\ \ \ ,
H^{\bullet}_{\ell ,\ell +1}=
\left.C\ \frac{d\tilde{R}^{\bullet}_{\ell ,\ell +1}}
{d\lambda}\right|_{\lambda =0}P
\end{equation}
with $\bullet=$A,B,C or D, which has the same
form as the
Hamiltonian of a spin-$S$ system. The dependence on
$\tilde{R}$ is
expected since this
matrix controls the non-cocommutativity of the coproduct,
which is the
relevant element for the construction of the Hamiltonian.
Furthermore, the dependence of (42)
on $\tilde{R}$ makes evident
the conclusions expressed in (41)
by just looking
at the relations for the various $\tilde{R}$ in (33-36).

In the cases B and D, using the explicit expression
\begin{equation}
R^D=\left(
\begin{array}{llll}
a & 0 & 0 & 0\\
0 & tc & b & 0\\
0 & b & t^{-1}c & 0\\
0 & 0 & 0 & a
\end{array}
\right)
\end{equation}
and eq. (15) (or (42)) we get \cite{mon,ku,aba}
\begin{equation}
H^D_{\ell ,\ell +1}=
g^2_{\ell +1}H^A_{\ell ,\ell +1}g^{-2}_{\ell +1}=
(t^{-1}S^+_{\ell}S^-_{\ell +1}+tS^-_{\ell}S^+_{\ell +1}+
2\cos\gamma S^3_{\ell}S^3_{\ell +1})\ ,
\end{equation}
apart from an additive constant. The similarity
transformation
generated by
\begin{equation}
K=
\exp \left[i\alpha \sum^{N}_{\ell
=1}\ (\ell -1)S^3_{\ell}\right]\ ,
\end{equation}
with $t=e^{i\alpha}$, takes $H^D$ to
$\tilde{H}^D=K^{-1}H^DK$, that is \cite{ku},
\begin{eqnarray}
\tilde{H}^D &=& \sum^{N-1}_{\ell =1}(S^+_{\ell}S^-_{\ell +1}+
S^-_{\ell}S^+_{\ell +1}+2\cos \gamma S^3_{\ell}S^3_{\ell +1}+
2\cos \gamma S^3_NS^3_1+ \\
&+&  t^{-N}S^+_NS^-_1+t^NS^-_NS^+_1)\nonumber
\end{eqnarray}
which is the well-known Hamiltonian for the XXZ
chain with twisted
boundary conditions $(S^{\pm}_{N+1}
\equiv t^{\pm N}S^{\pm}_1)$ \cite{alca}.

In summary, the Hamiltonians for the cases A,B,C and D are:
\begin{eqnarray}
H^C &=& H^A=H_{XXZ} \\
H^D &=& H^B=H^{twisted}_{XXZ}\ . \nonumber
\end{eqnarray}

Let us now consider the underlying algebraic
structure in  cases B,C and  D.
For the {\bf case B} the $L$ matrix is given by
$L^B_{1\ell}=g_1L^A_{1\ell}g_1$ with:
\begin{equation}
g_1=t^{S^3\otimes Z_{\ell}}=\left(
\begin{array}{ll}
t^{\frac{1}{2}\ Z_{\ell}} & 0 \\
0 & t^{-\frac{1}{2}\ Z_{\ell}}
\end{array}
\right)
\end{equation}
which leads, after factoring out the spectral
parameter, to
\begin{equation}
L^B_+=\left(
\begin{array}{ll}
q^{S^3}t^Z & \Omega S^-\\
0& q^{S^3}t^{-Z}
\end{array}
\right) \ \ \ ,
\ \ \ L^B_-=\left(
\begin{array}{ll}
q^{-S^3}t^Z & 0\\
-\Omega S^+ & q^{S^3}t^{-Z}
\end{array}
\right)\ ,
\end{equation}
where $Z$ as well as $\vec{S}$ are taken in an arbitrary
representation. The $R$-matrix is
the same as in case A and thus the equations
\begin{eqnarray}
R^B_{12}L^{\pm B}_{1\ell}L^{\pm B}_{2\ell} &=&
L^{\pm B}_{2\ell}L^{\pm B}_{1\ell}R^B_{12}\\
R^B_{12}L^{+B}_{1\ell}L^{-B}_{2\ell} &=&
L^{-B}_{2\ell}L^{+B}_{1\ell}R^B_{12}\nonumber
\end{eqnarray}
reduce to

\vspace{0.5cm}
\noindent
{\bf CASE B:}
\begin{eqnarray}
{[}S^3,S^{\pm}] &=& \pm S^{\pm}\nonumber \\
{[}S^+,S^-] &=& [2S^3]_q \\
{[}Z,\vec{S}] &=& 0 \ .\nonumber
\end{eqnarray}
The co-structure is, as usual, obtained by considering
the product of
two $L^B$ acting on two internal spaces (see eq. (22));
we find

\vspace{0.5cm}
\noindent
{\bf CASE B:}
\begin{eqnarray}
\Delta S^3 &=
& S^3\otimes {\reell}+{\reell}\otimes S^3 \nonumber \\
\Delta Z &=& Z\otimes {\reell}+{\reell}\otimes Z\\
\Delta S^{\pm} &=
& q^{S^3}t^{\mp Z}\otimes S^{\pm}+S^{\pm}\otimes
q^{-S^3}t^{\pm Z}\ ,\nonumber
\end{eqnarray}
that is, we have $SL_q(2)$ with a central element $t^Z$
\cite{bur,cha}. Notice that this central element
appears in a
non-trivial way in the co-product.

In  {\bf case C}, we have
$L^C_{1\ell}=g^{-1}_{\ell}L^A_{1\ell}g^{-1}_{\ell}$
with
$g_{\ell}=t^{Z\otimes S^3_{\ell}}=
1\!\!\!1\otimes t^{\frac{1}{2}\
S^3_{\ell}}$ which, after factoring out the spectral
parameter, gives.
\begin{equation}
L^C_+=\left(
\begin{array}{ll}
(qt^{-1})^{S_3} & \Omega S^-\\
0 & (qt)^{-S_3}
\end{array}
\right)\ \ \ ,
L^C_-=\left(
\begin{array}{ll}
(qt)^{-S_3} & 0\\
-\Omega S^+ & (qt^{-1})^{S_3}
\end{array}
\right)\ ,
\end{equation}
and
\begin{equation}
R^C=FR^AF=\left(
\begin{array}{rrrr}
q & 0 & 0 & 0\\
0 & t & \Omega & 0\\
0 & 0 & t^{-1} & 0\\
0 & 0 & 0 & q
\end{array}
\right)\ .
\end{equation}
Thus, by a similar procedure as in the previous
case we have for the algebra

\vspace{0.5cm}
\noindent
{\bf CASE C:}
\begin{eqnarray}
{[}S^3,S^{\pm}] &=& \pm S^{\pm} \\
t^{-1}S^+S^--tS^-S^+ &=& t^{-2S_3}[2S_3]_q\ ,\nonumber
\end{eqnarray}
and for the co-product

\vspace{0.5cm}
\noindent
{\bf CASE C:}
\begin{eqnarray}
\Delta S^3 &=
& S^3\otimes {\reell}+{\reell} \otimes S^3 \\
\Delta S^{\pm} &=
& (qt^{-1})^{S^3}\otimes S^{\pm}+ S^{\pm} \otimes (qt)^{-S^3}\
,\nonumber
\end{eqnarray}
which is the two-parameter deformed
$SL_{q,t}(2)$ algebra
\cite{li,sud,vat}. As noted in ref. \cite{cha}
the structure and
costructure of $SL_{q,t}(2)$ given respectively in
(55) and (56), are
actually isomorphic to $SL_q(R)$
(21,23) as one can easily check by defining
\begin{equation}
\tilde{S}_3=S_3\quad , \quad \tilde{S}^{\pm}=
S^{\pm}t^{S_3}
\end{equation}

The {\bf case D} follows similarly. In this case we
have $L^D_{1\ell}=
g^{-1}_{\ell}L^B_{1\ell}g^{-1}_{\ell}$ which
reduces to:
\begin{equation}
L^D_+=\left(
\begin{array}{ll}
(qt^{-1})^{S^3}t^Z & \Omega S^-\\
0 & (qt)^{-S^3}t^{-Z}
\end{array}
\right)\ \ \ ,
\ \ \ L^D_-=\left(
\begin{array}{ll}
(qt)^{-S^3}t^Z & 0\\
-\Omega S^+ & (qt^{-1})^{S^3}t^{-Z}
\end{array}
\right)\ ,
\end{equation}
and $R^D=R^C$. With this in hand, as in
the previous cases we have

\vspace{0.5cm}
\noindent
{\bf CASE D:}
\begin{equation}
{[}S^3,S^{\pm}]=\pm S^{\pm}\ \ ; \ \
t^{-1}S^+S^--tS^-S^+=t^{-2S^3}[2S^3]_q\ ; \
[Z,\vec{S}]=0 \ ,
\end{equation}
and for the co-product:

\vspace{0.5cm}
\noindent
{\bf CASE D:}
\begin{eqnarray}
\Delta S^3 &=
& {\reell} \otimes S^3+S^3\otimes {\reell} \nonumber \\
\Delta Z &=& Z\otimes {\reell} +{\reell} \otimes Z \\
\Delta S^{\pm} &=
& (qt^{-1})^{S^3}t^{\mp Z}\otimes S^{\pm}+S^{\pm}\otimes
(qt)^{-S^3}t^{\pm Z}\ ,\nonumber
\end{eqnarray}
which corresponds to the $SL_{q,t}(2)$ algebra
with a central element
$t^Z$; once more the central element appears in a
non trivial way in
the co-product.

By the transformation (57) eqs. (59-60) are sent
into the case B eqs. (51-52). By comparing the
cases related by a ``De Vega
transformation'' $(A\rightarrow B$ and
$C\rightarrow D$) one easily
realizes that the identities
\begin{eqnarray}
\Delta^B &=& F\Delta^AF^{-1}\\
\Delta^D &=& F\Delta^CF^{-1}\nonumber
\end{eqnarray}
hold and that the commutation relations do
not change, in agreement with
the Reshetikhin theorem \cite{li}, apart
from the enlargement of
the algebra due to the introduction of the
central element $Z$.

The Bethe ansatz equations can be easily
computed and are
those of the periodic XXZ chain in cases
$A,C$ and those of the
twisted XXZ chain in cases $B$ and $D$.

In summary, we have analysed three different
XXZ like models. They
are obtained by performing De Vega and dual
Reshetikhin
transformations, which are dual to each other,
in the $L$ and $R$
matrices of the periodic spin-$1/2$ chain. We have
shown that De Vega
transformation introduces a central element, $t^Z$,
in the underlying
algebraic
structure which gives rise to twisted
boundary conditions,
$S^{\pm}_{N+1}=t^{\pm N}S^{\pm}_1$. These cases
were here called  B
and D.
In cases $A$ and $C$ the is no central element in
the underlying algebraic structure and these cases
correspond to the
periodic spin-1/2 XXZ chain. As the only
consequence of the dual
Reshetikhin transformation is the
introduction of a
second parameter in the commutation relations,
without the presence
of
a non-trivial central element in the algebra,
it does not change the
Hamiltonian of the theory.

Moreover we have shown that in all the cases
we have considered, the
Hamiltonian can be written in an universal
way in terms of
$\tilde{R}$ (41), which, in our notation, is
not the $R$ matrix
entering into $FCR\ (RLL=LLR)$ but its counterpart
in the dual algebra.

Finally, we remark that the present discussion
can be generalized to
any model with invariance properties analogous
to eq. 30.


\newpage
\begin{thebibliography}{20}
\bibitem{vega} E. Lieb, Phys. Rev. Lett.
{\bf 18}, (1967) 692; Phys.
Rev. {\bf 162}, (1968) 162; Phys. Rev. Lett.
{\bf 18}, (1967) 1046;
Phys. Rev. Lett. {\bf 19}, (1967) 108.
\bibitem{re} H.J. De Vega, Nucl. Phys. {\bf B240}
(1984) 495; Int.
Jour. Mod. Phys. {\bf A4} (1989) 2371;
H.J. De Vega and E. Lopes,
Phys. Lett. {\bf B186} (1987) 180.
\bibitem{li} N. Reshetikhin, Lett. Math. Phys.
{\bf 20}, (1990) 331.
\bibitem{bax} R. Baxter, Ann. of Phys.
{\bf 70} (1972) 193; Ann. of
Phys. {\bf 70} (1972) 323.
\bibitem{yan} C.N. Yang, Phys. Rev. Lett.
{\bf 19} (1967) 1312.
\bibitem{lya} E. Sklyanin, L. Takhtajan and
L. Faddeev, Teor. Matem.
Fiz. {\bf 40} (1979) 194; L. Takhtajan and
L.Faddeev, Russian Math.
Surveys {\bf 34} (1979) 11; N. Reshetikhin,
L. Takhtajan and L.
Faddeev, Leningrad Math. J. {\bf 1} (1990) 193.
\bibitem{fil} L. Faddeev, in ``New Problems,
Methods and
Techniques in Quantum Fielf Theory and
Statistical Mechanics'', edited
by M. Rasetti, World Scientific Pub. (1990);
 ``Algebraic Aspects of
Bethe-Ansatz'', preprint ITP-SB-94-11.
\bibitem{for} For a review see: ``Quantum Inverse
Scattering Method
and Correlation Functions'' by V.E. Korepin,
N.M. Bogoliubov and A.G.
Izergin, Cambridge U.P. (1993).
\bibitem{ku} M.R-Monteiro, I. Roditi,
L.M.C.S. Rodrigues and S.
Sciuto, ``The Quantum Algebraic Structure
of the Twisted XXZ Chain'',
preprint CBPF-NF-054/94 and DFTT-045/94,
HEP-TH/9410144, to appear in
Mod. Phys. Lett. A.
\bibitem{drin} P. Kulish and N. Reshetikhin,
Zap. Nauch. Seminarov LOMI
{\bf 101}, (1981) 101; J. Sov. Math. {\bf 23}
(1983) 2435.
\bibitem{la} V. Drinfeld, Sov. Math. Dokl.
(1985) 254; M. Jimbo,
Lett. Math. Phys. {\bf 10}, (1985) 63; {\bf 11}
(1986) 247.
\bibitem{mon} L. Hlavat\'y, ``Solution to the
YBE Corresponding to the
XXZ Models in an External Magnetic Field'', preprint
E5-85-959, Dubna
(1985); J. Math. Phys. {\bf A 27} (1994) 5645.
\bibitem{aba} J. Abad and M. Rios, ``Integrable Spin
Chain Associated
to $\widehat{SL}_q(n)$ and $\widehat{SL}_{p,q}(n)$'',
preprint DFTUZ/94/21, HEP-TH/9410193.
\bibitem{alca} F. Alcaraz, U. Grimm and V. Rittenberg,
Nucl Phys.
{\bf B316} (1989) 735; V. Pasquier and H. Saleur,
Nucl. Phys. {\bf B330},
(1990), 523.
\bibitem{bur} C. Burdik and P. Hellinger, J. Phys.
{\bf A25} (1992) L629.
\bibitem{cha} R. Chakrabarti and R. Jagannathan,
J. Phys. {\bf A27}
(1994) 2023.
\bibitem{sud} A. Sudbery, J. Phys. {\bf A23},
(1990) L697; M.
Takeuchi, Proc. Jap. Acad. {\bf 66}, (1990) 112.
\bibitem{vat} C. Burdik and L. Hlavat\'y, J. Phys.
{\bf A24}, (1991)
L165.
\end{thebibliography}
\end{document}